\documentstyle[twoside,fleqn,espcrc2,epsfig]{article}

\newcommand{\AmS}{{\protect\the\textfont2
  A\kern-.1667em\lower.5ex\hbox{M}\kern-.125emS}}

\title{Deep-inelastic Electron-Photon Scattering at High $Q^{2}$ : Neutral and
Charged Current Reactions}

\author{ A.~Gehrmann--De Ridder \address{
Institut f\"{u}r Theoretische  Teilchenphysik, Universit\"{a}t
Karlsruhe, D-76128 Karlsruhe, Germany and \\
Deutsches Elektronen-Synchrotron DESY, D-22603
  Hamburg, Germany.}}

\begin{document}



\begin{abstract}
We present the results of a calculation 
of deep inelastic electron-photon scattering at a linear collider for very
high virtuality of the intermediate gauge boson up to NLO in
perturbative QCD. 
The real photon is 
produced unpolarized via the Compton back scattering
of laser light of the incoming beam.
For $Q^2$ values close to the masses squared of the Z and W gauge bosons,
the deep inelastic electron-photon scattering process 
receives important contributions
not only from virtual photon exchange but also from the exchange of a Z-boson 
and a W-boson.
We find that the total cross section for center of mass energies above
$500 \rm{GeV}$ is at least of ${\cal O}(pb)$ 
and has an important charged current contribution.
\end{abstract}                                                                

\maketitle

\vspace*{-8.5cm} \noindent TTP99-30
\vspace*{ 7.2cm}


\section{Introduction}

Processes induced by initial state photons provide us with an
interesting testing ground for QCD. 
As a photon  can interacts directly through a pointlike coupling with quarks 
or through its parton content like a hadron 
it has a twofold nature.
Its point-like interaction gives rise to perturbatively calculable
short-distance contributions  \cite{phostruc}
while its hadron-like or resolved part  
cannot be described with perturbative
methods. It is described in terms of the parton distribution functions
inside the photon. These parton distributions obey a perturbative
evolution equation with a non-perturbative boundary condition usually
parameterized in the form of an initial distribution at some low starting
scale $\mu_0$. 
The pointlike and resolved processes contribute to the various real
structure functions entering the cross section for 
$e^+e^- \to e^- + \gamma \to l +X$, the object of this study. 
 
\section{The photon spectrum}
The process of Compton backscattering of laser beams 
off highly energetic electrons/positrons offers an efficient mechanism 
to transfer a large fraction of the lepton energy to the photon 
which structure can be investigated in deep inelastic electron-photon 
scattering.

The differential 
Compton cross section for the process $e^{\pm} +\gamma \to e^{\pm}
+\gamma^{'}$, where the polarization of the final state photon 
$\gamma^{'}$ is not observed takes the following form \cite{kuehn,ginzburg},
\begin{eqnarray} 
\frac{{\rm d}\sigma}{{\rm d}y_{\gamma}}&=&
\frac{\pi \alpha^2}{x_{0} m^2_{e}}\;\ \Big \{ \frac{1}{1-y}
+(1-y)-4r(1-r) 
\nonumber \\
& & + P_{e}P_{\gamma}\,x_{0}\,r\,(1-2r)(2-y) \Big \}.
\end{eqnarray}
The energy transferred from the electron to the
backscattered photon is denoted by $y_{\gamma}$. $P_{e}, P_{\gamma}$ are
the helicities of the incoming lepton and laser photon with 
$-1\leq P_{e},P_{\gamma} \leq 1$.
In the above formula, the ratio $r=y/[(1-y)x_0]$ while the fractional 
energy of the final state photon $y_{\gamma}$ is bounded by
$y_{\gamma}\leq \frac{x_0}{1+x_0}$. 
The parameter $x_0$ is defined as $\frac{4 E w_{0}}{m_e^2}$ where 
$E$ and $w_0$ are the energies of the incoming electron and photon.
By tuning the energy of the incoming photon, the parameter 
$x_{0}$ can be chosen close to 4.83, just below the threshold for
production of $e^+e^-$ pairs from the collision of laser and final
state photons. To give an order of magnitude, for an energy of the
incoming lepton of $250\; \rm{GeV}$, the laser energy is of the order of
$1\; \rm{eV}$.   

It is worth noting that the photon luminosity  
for backscattered photons 
is of the same order as the initial electron luminosity. 
It is therefore enhanced  
compared to the photon luminosity obtained for
photon produced by bremsstrahlung off the lepton which is
of ${\cal O}(\alpha)$.
Furthermore, by choosing the helicities of laser and electron beams 
to be opposite, the photon spectrum is peaked at high energies. 
In this region, about 80\% of the energy of the electron/positron beam 
can be transferred to the backscattered photon.
In the following we shall give our predictions for the most 
advantageous photon spectrum and will therefore consider the
backscattered photon obtained for opposite 
polarizations of electron and laser photon, with $x_0=4.83$.
Our results will be obtained for the cross section 
$e^+e^- \to \l +X$  which is related to the cross section for 
$ e^- \gamma \to l+X$ as follows,
\begin{equation}
{\rm d}\sigma_{e^+e^- \to \l +X}=\int {\rm d}y_{\gamma}
\frac{1}{\sigma_c}\frac{{\rm d}\sigma_c}{{\rm d}y_{\gamma}}
{\rm d}\sigma_{e^+ \gamma \to \l +X}.
\end{equation}

\section{Deep-Inelastic Electron-Photon scattering} 

At momentum transfer squared $Q^2$ close to the Z and W masses squared, 
the deep inelastic cross section for $e +\gamma \to l +X$ becomes sensitive to
contributions from these exchanges \cite{cordier}.
Neutral current deep inelastic events are characterized
by the exchange of a photon or a Z boson, and have an electron and a jet
(or jets) in the final state. Charged current events arise
via the exchange of a W-boson and contain an
undetected neutrino and a jet (or jets) in the final state.
The presence of an neutrino is as usual detected as missing transverse
momentum.
Having these signatures, those processes constitute potential background
sources for searches for new physics at the future $e^+e^-$ linear collider. 
To give a prediction for the size of these contributions is
precisely the aim of this study.

\subsection{Kinematics}

The neutral current differential cross section is parameterized by the 
$x$ and $Q^2$ dependent structure
functions $F^{\gamma,NC}_{2}$ and $F^{\gamma,NC}_{L}$,
\begin{eqnarray}
\label{eq:NC}
\lefteqn{
\frac{{\rm d}^2 \sigma^{NC}(e +\gamma \to e +X)}{{\rm d}x {\rm d}Q^2} =}
\nonumber \\
& &\frac{2 \pi \alpha^2}{x Q^4}\left[
(1+(1-y)^2)F^{\gamma,NC}_{2}
-y^2 F^{\gamma,NC}_{L}\right]
\end{eqnarray}
while in the charged current differential cross section enter  
the $x$ and $Q^2$ dependent structure
functions $F^{\gamma,CC}_{2}$, $F^{\gamma,CC}_{L}$ and $F^{\gamma,CC}_{3}$
multiplied by weak propagator and coupling factors as follows 
\begin{eqnarray}
\label{eq:CC}
\lefteqn{
\frac{{\rm d}^2 \sigma^{CC}(e^{\pm} +\gamma  \to 
\stackrel{{\scriptscriptstyle (-)}}{\nu} +X)}{{\rm d}x {\rm d}Q^2} =}
\nonumber \\ 
& &\frac{1}{4}\frac{\pi \alpha^2}{x Q^4}\; \frac{Q^4}{[Q^2 +M^2_{W}]^2}\; 
\frac{(1\pm P)}{\sin^4 \Theta_w} \nonumber \\
& &\left[(1+(1-y)^2)F^{\gamma,CC}_{2} \right.
\nonumber \\
& & \left. \quad \mp (1-(1-y)^2)x F^{\gamma,CC}_{3} 
-y^2 F^{\gamma,CC}_{L}\right].
\end{eqnarray}
$M_{W}$ and $\Theta_{W}$ are the W-mass and the weak Weinberg angle.
P denotes the degree of left-handed longitudinal polarization ($P=-1$
for left-handed electrons, $P=1$ for right-handed electrons).

The kinematics of these neutral and charged current reactions is
described by the Bjorken variables $x$ and $y$. 
Those can be expressed in terms of the 
four-momentum transfer, $Q^2$, 
the hadronic energy $W$ , the lepton energies and
scattering angles. Indeed we have, 
\begin{eqnarray}
x &=& \frac{Q^2}{2 q. p_{\gamma}}=\frac{Q^2}{Q^2 +W^2} \nonumber \\
y &=& \frac{q. p_{\gamma}}{k. p_{\gamma}}=1-\frac{E}{E'}\cos^2(\Theta/2).
\end{eqnarray}
The initial state photon is real and on-shell and  
the lepton masses can be set to zero.

\begin{figure}[t]
\vspace{-7mm}
\begin{center}
~\epsfig{file=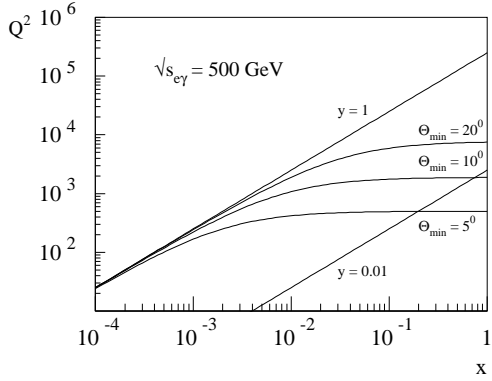,angle=-90,width=7cm}
\end{center}
\vspace{-7mm}
\caption{The kinematical region corresponding to $\Theta_{min}=5, 10,
20$ degrees, for $\sqrt s_{e \gamma}=500 \; \rm{GeV}$.} 
\label{fig:kine}
\end{figure}

Experimentally, one needs to constrain the lepton scattering angle
$\Theta_e$ to be larger than some minimum angle $\Theta_{min}$.
Equivalently, one can also require, 
\begin{eqnarray}
\label{eq:Q2}
Q^2 & > & \frac{4 E^2 (1-c_o) x s_{e \gamma}}{x s_{e \gamma} (1+c_o) +4 E^2 (1-c_o)}
\nonumber \\
Q^2 & > & Q^2(x,s_{e \gamma},c_o)
\end{eqnarray}
with $c_o=\cos \Theta_{min}$ and $s_{e \gamma}$ the electron-photon
center-of-mass energy squared.
This constraint on the 4-momentum transfer $Q^2$ is shown as a function of $x$
in Fig.~\ref{fig:kine} for $\sqrt s_{e \gamma}=500\; \rm{GeV}$and 
for $\Theta_{min}=5, 10, 20$ degrees. 
The corresponding allowed kinematical regions are then
situated between  $Q^2=Q^2(x,s_{e \gamma},c_o)$ and $Q^2=xs_{e \gamma}$ (corresponding
to $y=1$).

In this study, we will  choose to present our results 
for fixed minimum values of $Q^2$. For illustrative purpose, in Table
\ref{table:1} we present the choices of $Q^2_{min}$ corresponding to different
electron-photon center of mass energy and $\Theta_{min}$ values.

\subsection{Neutral and charged current structure functions}

The neutral and charged current structure functions appearing 
in eqs.~(\ref{eq:NC}) and (\ref{eq:CC}) depend on the Bjorken  
variable $x$ and on the 4-momentum transfer $Q^2$; in our study, 
$x$ typically lies between $10^{-3}$ and 0.8, while $Q^2$ takes values
close to the masses squared of the Z and W bosons. $Q^2$ being the only 
large scale in this study, the structure functions follow a
perturbative expansion in $\alpha_s \log(Q^2)$, 
an expansion in the leading and next-to-leading logarithmic terms. 
Furthermore, neutral and charged structure 
functions have different components coming from the various 
flavour-dependent parton distributions building them up.
\begin{table}[t]
\caption{The different $Q^2_{min}$values, see text}  
\label{table:1}
\vspace{0.3cm}
\begin{tabular}{ |c|c c c||c|}
\hline
{\footnotesize $\sqrt s_{e \gamma} \;(\rm{GeV})$}  &500 & 1000 & 2000 &   \\  \hline
$Q^2_{min}$ & $10^4$ & $5.10^4$ & $10^5$ & $20^{\circ}$ \\ \cline{2-5}
$(\rm{GeV}^2)$   & $10^3$ & $5.10^3$ & $5.10^4$ & $10^{\circ}$\\ \cline{2-5}
& $5.10^2$ & $5.10^3$ & $10^4$ & $5^{\circ}$ \\ \hline 
\end{tabular}
\end{table}

The neutral current structure functions $F^{\gamma,NC}_{2}$ and  
$F^{\gamma,NC}_{L}$ receive contributions from 6 quark
flavours. Except for the top quark, the masses of the 5 lightest quark
are negligible compared to the scale $Q^2$. These flavours can be treated
as massless and their contributions  to $F^{\gamma,NC}_{2}$ and  
$F^{\gamma,NC}_{L}$ are connected to the massless parton distributions 
in the photon. Those distributions are solution 
of the all-order DGLAP evolution
equations in the leading or next-to-leading logarithmic approximation
where respectively only terms in $\alpha^n_s \log^{n+1}Q^2$ or 
$\alpha^n_s \log^{n+1}Q^2$ and  $\alpha^n_s \log^{n}Q^2$ are retained.

The different quark contributions need furthermore to be multiplied 
by the corresponding quark charges.
As the virtual $\gamma$ exchange is supplemented by Z-boson exchange at
high virtuality $Q^2$  the charge factor becomes,
\begin{eqnarray}
\lefteqn{ e^2_q \to \hat{e}^2_{q}:= }       \nonumber \\
 &\frac{1}{4}\sum_{i,j=L,R} \left [ e_q 
-\frac{Q^2}{Q^2 +M_{Z}^2}\frac{Z_i(e)Z_j(q)}{\sin^2\Theta_w
  \cos^2\Theta_w}\right] ^2
\end{eqnarray}
where the electroweak Z charges are given by 
\begin{eqnarray}
Z_L(f)&=& I_{3L}(f) -e_f \sin^2\Theta_w
\nonumber \\
Z_R(f)&=& -e_f \sin^2\Theta_w 
\end{eqnarray}
for the left and right-handed Z-couplings. 
Only the 4 lightest quark flavours contribute to the charged
current structure functions through a combination
of massless quark $q(x,Q^2)$ and antiquark $\bar{q}(x,Q^2)$ 
distributions in the photon.
Bottom and top contributions are here neglected as these are considerably
smaller.

More explicitly, at the lowest order,
for the neutral current structure
functions we have,
\begin{eqnarray}
\label{eq:f2lo}
F^{\gamma,NC}_{2}&=& x\sum_{q} \hat{e}^2_{q}
\left \{ q(x,Q^2) +\bar q(x,Q^2) \right \}
\end{eqnarray}
where the sum runs over the 5 lightest flavours.
The top quark contribution to $F^{\gamma,NC}_{2}$ is given at this
order by 
\begin{equation}
F^{\gamma,NC}_{2,t}=
\frac{\alpha_s(Q^2)}{\pi} \left [ C_{g} \otimes g(x) \right ] 
+ \frac{\alpha}{\pi}\hat{e}^2_{q}\;  C_{\gamma}.
\end{equation}
$F^{\gamma,NC}_{L,t}$ is the only non-vanishing longitudinal 
structure function at this order. 
  
For the scattering of an $e^-$ with the photon, 
the charged current structure functions $F^{\gamma,CC}_{2}$ and 
$F^{\gamma,CC}_{3}$ at the lowest order are respectively given by,
\begin{eqnarray}
\label{eq:f3lo}
F^{\gamma,CC}_{2}&=& x \left[u(x,Q^2) +c(x,Q^2) 
\right. \nonumber\\
& & \left. + \bar{d}(x,Q^2) +\bar{s}(x,Q^2) \right] \nonumber \\
F^{\gamma,CC}_{3}&=& \left[u(x,Q^2) +c(x,Q^2) \right. 
\nonumber\\
& & \left.-\bar{d}(x,Q^2) -\bar{s}(x,Q^2) \right]. 
\end{eqnarray}
For $e^+$ scattering, one has to make the following 
replacements: $u(x,Q^2) +c(x,Q^2) 
\to \bar{u}(x,Q^2) +\bar{c}(x,Q^2)$ and $\bar{d}(x,Q^2)
+\bar{s}(x,Q^2) \to d(x,Q^2) + s(x,Q^2)$.
The CKM matrix is approximated by the unity matrix, flavour mixing
effects can here be neglected.

At the next-to-leading order, corrections proportional to $\alpha_s \log(Q^2)$ 
have to be included in all the above leading-order expressions. To the
terms proportional to the parton distributions themselves, terms
involving convolution of these with quark ($C_{q}$) and gluon ($C_g$) 
coefficient functions have to be taken into account.
For example, for the light flavours, 
$F^{\gamma,NC}_{2}$ given at leading order in
eq.(~\ref{eq:f2lo}) becomes in the $\overline{\rm MS}$ scheme 
at the next-to-leading order 
\begin{eqnarray}
F^{\gamma,NC}_{2}= x\sum_{q}  \hat{e}^2_{q}
\left \{ q(x,Q^2) +\bar q(x,Q^2) \right \}
\nonumber \\
\quad +\frac{\alpha_s(Q^2)}{2 \pi} \Big[ C_{q}\otimes 
\left \{ q(x,Q^2) +\bar q(x,Q^2) \right \}
\nonumber\\
\quad + C_{g}\otimes g(x,Q^2)\Big] +\frac{\alpha}{\pi} C_{\gamma,2}.
\end{eqnarray}

At this order, we use the 
beyond-leading-logarithmic (BLL) GRV \cite{grv} massless parton
distributions $q(x,Q^2)$ and $g(x,Q^2)$.  
These are given in the $DIS_{\gamma}$ factorization scheme, defined by,
\begin{equation}
q(x,Q^2)_{DIS_\gamma}=q(x,Q^2)_{\overline{\rm MS}} 
+\frac{\alpha}{2 \pi} C_{\gamma,2}.
\end{equation}
As a consequence, in this $DIS_{\gamma}$ factorization scheme
the direct term $C_{\gamma,2}$ is absent from the expression of
$F^{\gamma,NC}_{2}$. The gluon distribution $g(x,Q^2)$ remains  unaffected by 
the change of factorization scheme. 
The precise expressions of the other neutral and charged structure functions 
at the next-to-leading order will be given in \cite{gz}.

\section{Results and conclusion}

As mentioned before, we shall present our results for fixed
values of $Q^2_{min}$, namely $Q^2_{min}=10000\; \rm{GeV}^2$ and  
$Q^2_{min}=1000\; \rm{GeV}^2$. Since the leading and next-to-leading order
results are found to be very close to each other, all results 
will be given at the next-to-leading order level only. 
The NLO corrections being at most of the percent level indicates furthermore 
that the obtained results are perturbatively stable.  

\begin{figure}[htb]
\vspace{-7mm}
\begin{center}
~\epsfig{file=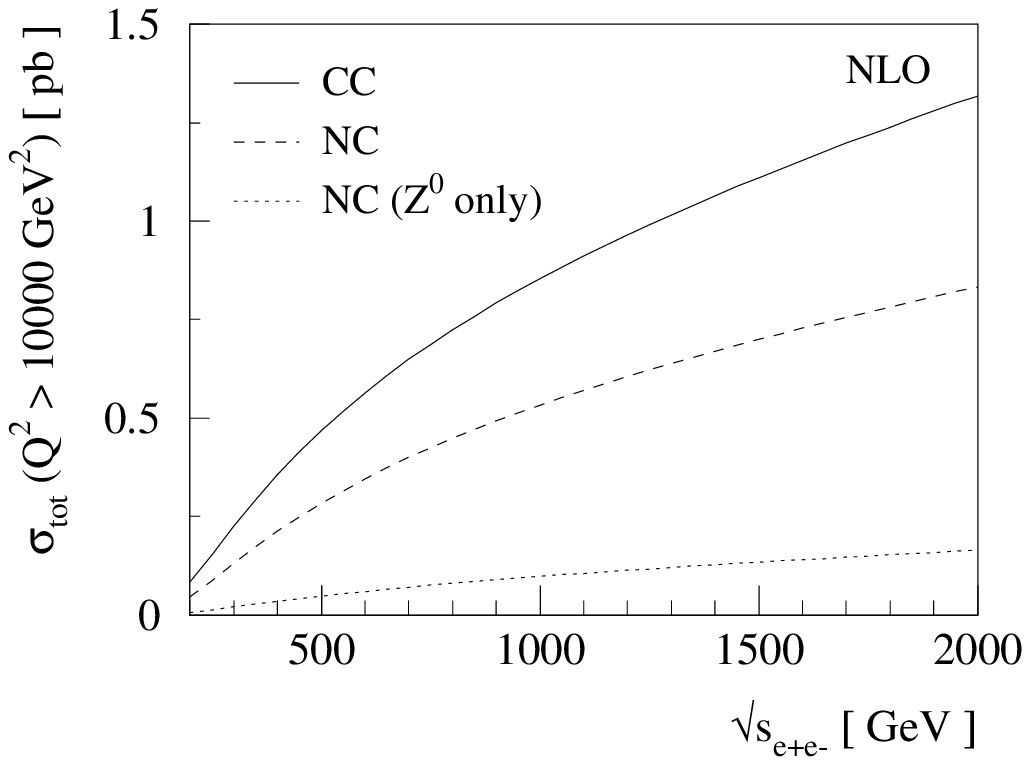,angle=-90,width=7cm}
\end{center}
\vspace{-7mm}
\caption{The total cross section as function of $\sqrt s$ for 
$Q^2_{min}=10000 \;\rm{GeV}^2$.}
\label{fig:7}

\begin{center}
~\epsfig{file=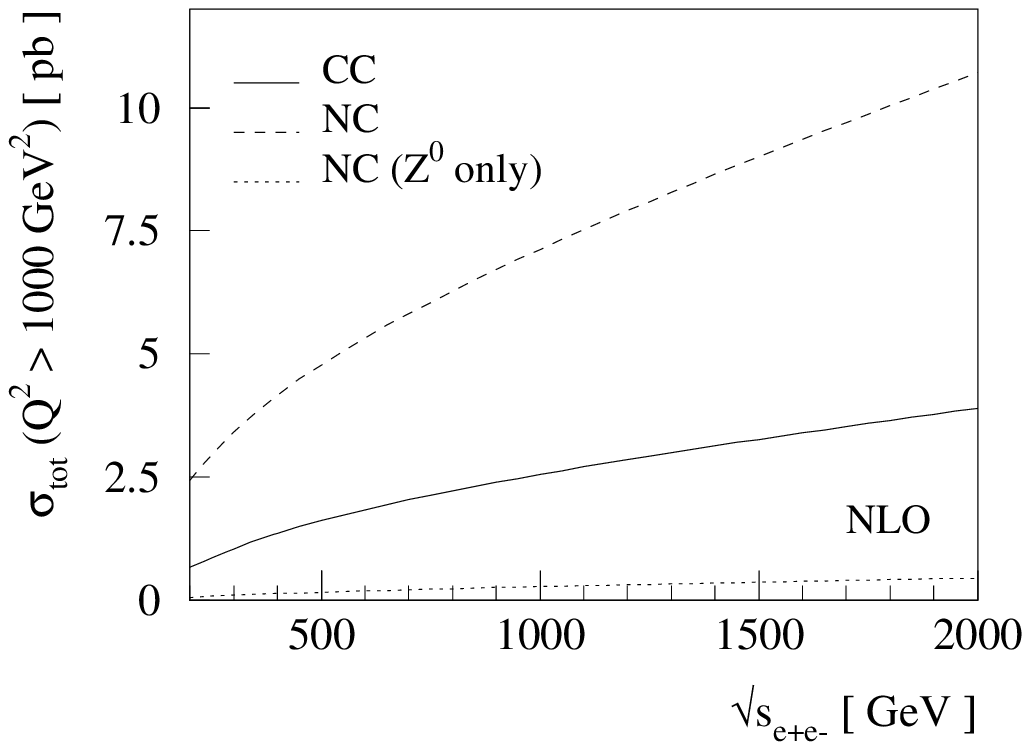,angle=-90,width=7cm}
\end{center}
\vspace{-7mm}
\caption{ The total cross section as function of $\sqrt s$ for 
$Q^2_{min}=1000 \;\rm{GeV}^2$.}
\label{fig:7b}
\end{figure}

In Figs.~\ref{fig:7} and ~\ref{fig:7b} we present the different
contributions to the total NC and CC cross sections as a function of the
electron-positron center of mass energy, the latter varying between 
$\sqrt s=200 \;\rm{GeV}$ and $\sqrt s=2000\; \rm{GeV}$ for virtualities of the
intermediate gauge boson equal to $10000 \;\rm{GeV}^2$ and  
$1000\; \rm{GeV}^2$ respectively.

The cross sections being inversely proportional to $Q^4$,
their value is dominated by the smallest $Q^2$ values.
As can be seen in these figures, for $Q^2_{min}=10 000 \;\rm{GeV}^2$, the
largest cross section, the total charged current cross section is of 
${\cal O}(pb)$ while for $Q^2_{min}=1000 \;\rm{GeV}^2$, the dominant 
neutral current cross section is about 10 times larger.
For this latter choice of $Q^2_{min}$ the charged current cross section is
approximately a third of the neutral current cross section.
By these high virtualities of the intermediate gauge boson, the exchange
of a W-boson gives always rise to significantly high cross sections. 
The contribution arising from the 
exchange of a Z-boson is the smallest for both choices of $Q^2_{min}$.

\begin{figure}[htb]
\vspace{-7mm}
\begin{center}
~\epsfig{file=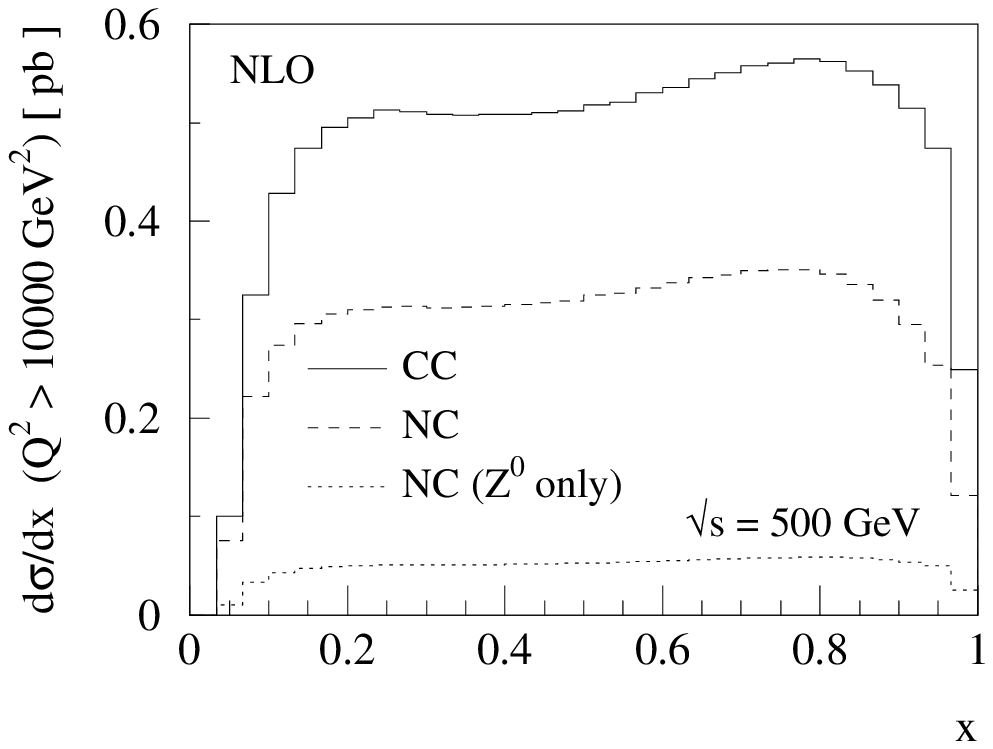,angle=-90,width=7cm}
\vspace{-7mm}
\caption{The differential cross section as a function of x for 
$Q^2_{min}=10000 \;\rm{GeV}^2$.}
\label{fig:8}
\end{center}

\vspace{-7mm}
\begin{center}
~\epsfig{file=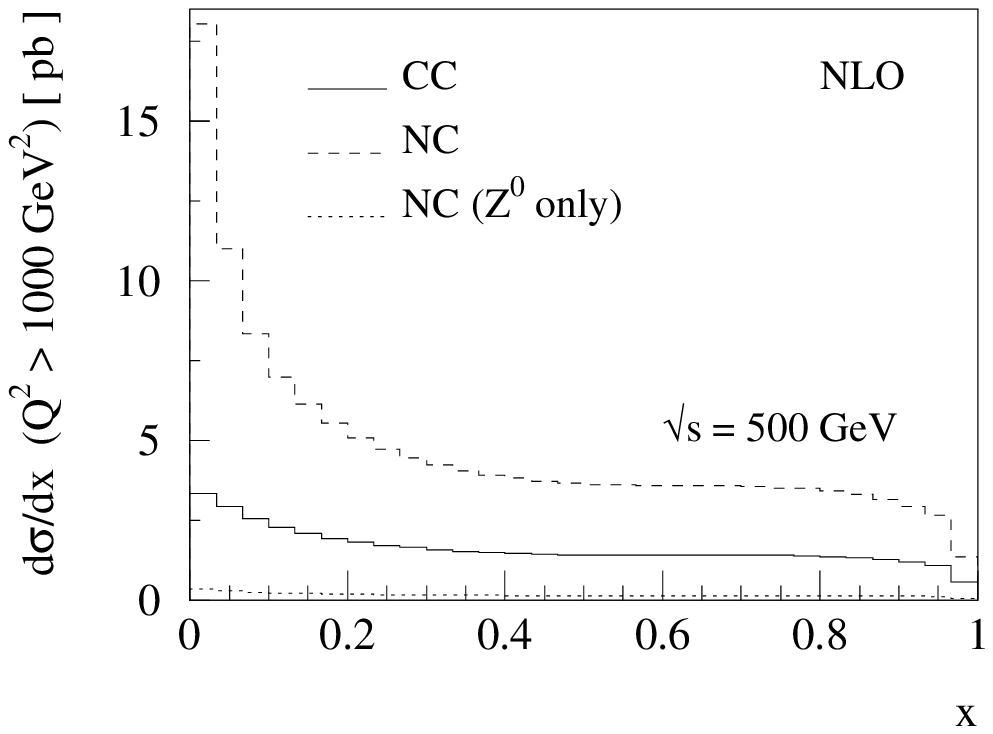,angle=-90,width=7cm}
\vspace{-7mm}
\caption{The differential cross section as a function of x for 
$Q^2_{min}=1000 \;\rm{GeV}^2$.}
\label{fig:8b}
\end{center}
\end{figure}

The differential cross sections with respect to $x$, 
which are shown for $\sqrt s=500 \;\rm{GeV}$ in Figs. ~\ref{fig:8} and
~\ref{fig:8b} differ not only in magnitude but also in shape.
This can be understood as follows.
For a given value of $x$, the allowed phase space regions in the 
$(x,Q^2)$ plane corresponding to the two values of $Q^2_{min}$ chosen
here, differ significantly from each other. 
As can be seen in Fig.~\ref{fig:kine}, this results in an enhanced
importance of the small $x$ region for smaller $Q^2_{min}$.

\begin{figure}[htb]
\vspace{-7mm}
\begin{center}
~\epsfig{file=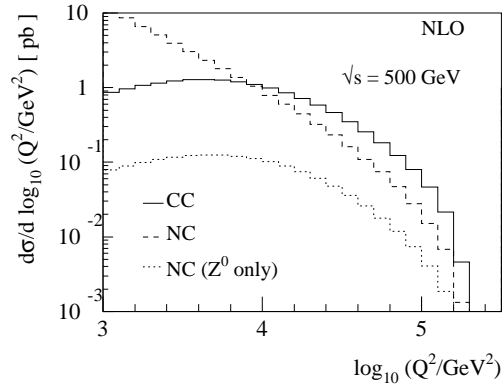,angle=-90,width=7cm}
\vspace{-7mm}
\caption{The differential cross section as a function of $Q^2$.}
\label{fig:9b}
\end{center}
\end{figure}

Finally, we also present the differential cross section with respect to
$Q^2$ for $Q^2$ values varying between $1000 \;\rm{GeV}^2$ and the
electron-positron center-of-mass energy squared $s$, for $\sqrt s=500 \;\rm{GeV}$.
As can be seen in 
Fig.~\ref{fig:9b} the neutral and charged current cross sections 
fall off as $Q^2$
increases. Above $Q^2=10000 \;\rm{GeV}^2$ the CC cross section is the
largest while below this value of $Q^2$ the NC reaction induced 
by the exchange of an intermediate photon
gives rise to  the largest differential cross section with respect to $Q^2$.

\section*{Acknowledgements}
I wish to thank H.~Spiesberger and P.~Zerwas for a fruitful
collaboration and A.~Wagner for financial support during my stay at DESY
where part of this work was carried out.
Furthermore I wish to thank S.~Soldner-Rembold for 
organizing an interesting and pleasant workshop.

\end{document}